\documentclass[prb,showpacs,twocolumn,floatfix]{revtex4}

\usepackage{graphicx} 
\usepackage{color} 
\usepackage{amsmath} 
\usepackage[urlcolor=blue, hyperindex, colorlinks, bookmarks=true,linkcolor=black,citecolor=black]{hyperref} 
\usepackage{amssymb}

\newcommand{\bra}[1]{\langle{#1}|}
\newcommand{\ket}[1]{|{#1}\rangle}

\newcommand{\mbf}[1]{\mathbf{#1}}

\begin{document}

\title{Entangled photons on demand: Erasing which-path information with sidebands}
\date{\today}
\author{W. A. Coish}
\affiliation{Institute for Quantum Computing and Department of Physics and Astronomy, University of Waterloo, Waterloo, Ontario N2L 3G1, Canada}
\author{J. M. Gambetta}
\affiliation{Institute for Quantum Computing and Department of Physics and Astronomy, University of Waterloo, Waterloo, Ontario N2L 3G1, Canada}	  	  

\begin{abstract}
The biexciton cascade in a quantum dot can be used to generate entangled-photon pairs rapidly and deterministically (on demand). However, due to a large fine-structure splitting between intermediate exciton energy levels, which-path information encoded in the frequencies of emitted photon pairs leads to a small degree of entanglement. Here we show that this information can be efficiently erased by modulating the exciton and biexciton energy levels, giving rise to new decay paths through additional sidebands.  The resulting degree of entanglement is substantial, and can be made maximal through spectral filtering, with only a nominal reduction in collection efficiency.
\end{abstract}


\maketitle

A fast and deterministic source of highly entangled photon pairs is a central requirement in schemes for measurement-based quantum computing \cite{KnillRaussendorf} and long-ranged quantum communication \cite{BriegelDur}.  One of the most promising methods for entangled-photon pair generation makes use of a biexciton cascade in semiconductor quantum dots \cite{Benson2000a}.  These systems offer the possibility of rapid and deterministic generation of entangled photon pairs, in contrast to current schemes based on parametric down conversion, which are inherently stochastic and have pair-generation rates limited by the creation of multiple pairs at high pump power.

Significant progress has been made in recent years toward the realization of an efficient source of on-demand entangled photon pairs \cite{Stevenson2006a,Akopian2006a,Young2009a}. However, the achievable degree of entanglement (or efficiency of pair collection) is severely limited by the electron-hole exchange interaction, giving rise to an intrinsic fine-structure splitting (FSS) separating the intermediate exciton states \cite{GammonBester, Stace2003a,Stevenson2006a,Akopian2006a,Avron2008a}.  Several methods have been proposed and used to eliminate the FSS, including applied dc electric \cite{Kowalik2005a,Gerardot2007a,Reimer2008a} and magnetic fields \cite{Stevenson2006a}, strain \cite{Seidl2006a}, the ac Stark effect \cite{Jundt2008a,Muller2008a}, and strong coupling to a cavity \cite{Stace2003a,JohnePathak}.  Additional schemes have been proposed and demonstrated that give entangled pairs in spite of the FSS, including post-selection based on frequency \cite{Akopian2006a,Meirom2008a}, as well as schemes involving `time reordering' of emitted photons \cite{Reimer2007a,Avron2008a,Troiani2008a}.  A further possibility is to anneal self-assembled dots after growth to reduce the FSS \cite{Langbein2004a}. There has been some success in creating highly-entangled photon pairs using dots that are specially selected for their small FSS \cite{Young2009a}, although this method requires the selection of one out of many dots in an ensemble. While each of these methods shows promise or a measured degree of success, various complications have made it difficult to generically create entangled photon pairs with both a high degree of entanglement, and a high efficiency of photon pair generation for typical quantum dots, where the FSS is large compared to the radiative linewidth.

Here, we introduce an alternative scheme for erasing `which-path' information in a biexciton cascade.  This scheme is conceptually and technologically simple, avoids pitfalls associated with methods using large dc electric fields, and works for typical quantum dots, with large associated FSS. In this scheme, sidebands can be generated through modulation of the biexciton and exciton energy levels.  By choosing the modulation frequency to coincide with the FSS, and by appropriately tuning the modulation amplitude, it is possible to recover a significant degree of entanglement for emitted photon pairs.  
\begin{figure}[t]
	\centering
		\includegraphics[width=0.45\textwidth]{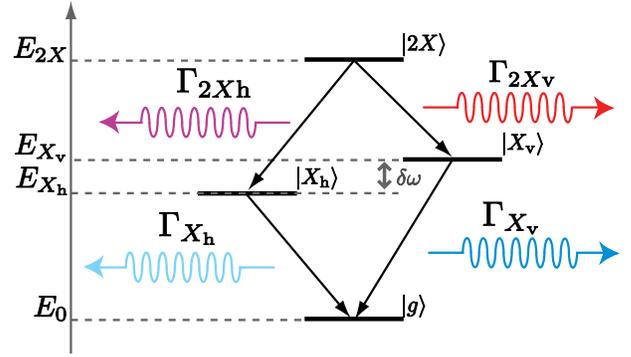}
	\caption{(Color online) Energy levels and decay rates for the biexciton cascade.  The exciton states $\ket{X_\mathrm{h}}$ ($\ket{X_\mathrm{v}}$) couple only to horizontally (vertically) polarized light.  Decay from the biexciton state $\ket{2X}$ to the ground state $\ket{g}$ results in the emission of two photons, which may be entangled in polarization. }
	\label{fig:Schematic}
\end{figure}

The canonical biexciton cascade is shown schematically in Fig. \ref{fig:Schematic}.  The two intermediate bright exciton states couple to two orthogonal linear polarizations. Within the rotating-wave approximation, the dot-light coupled system is then described by the Hamiltonian (we set $\hbar=1$):
\begin{equation}\label{eq:Hamiltonian}
		H=H_0+\sum_{\mathbf{k}\lambda} g_{\mathbf{k}\lambda}\sigma_{X_\lambda,2X}a_{\mathbf{k}\lambda}^\dagger
		+f_{\mathbf{k}\lambda}\sigma_{0,X_\lambda}a_{\mathbf{k}\lambda}^\dagger+\mathrm{h.c.},
\end{equation}
where $\sigma_{i,j}=\ket{i}\bra{j}$ describes an excitation on the dot, the noninteracting Hamiltonian is $H_0=\sum_{j}E_j\ket{j}\bra{j}+\sum_{\mathbf{k}\lambda}\nu_k a_{\mathbf{k}\lambda}^\dagger a_{\mathbf{k}\lambda}$, $j=\{g,X_\mathrm{h},X_\mathrm{v},2X\}$, and $\omega_{ij}=E_i-E_j$.  Here, $a_{\mbf{k}\lambda}^\dagger$ creates a photon with wavevector $\mathbf{k}$ and linear polarization $\lambda=\{\mathrm{v},\mathrm{h}\}$ having frequency $\nu_k$. The state $\ket{g}$ describes the empty dot (without excitons), $\ket{2X}$ describes the (non-degenerate) biexciton ground state, and $\ket{X_\mathrm{h,v}}$ denote the two bright exciton states, separated in energy by the FSS: $E_{X_\mathrm{v}}-E_{X_\mathrm{h}}=\delta\omega$.

The dot is initialized to the biexciton state with the state of the photon field given by vacuum ($\ket{\psi(0)}=\ket{2X}\otimes\ket{0}$).  This state evolves coherently under the action of the Hamiltonian (Eq. (\ref{eq:Hamiltonian})), giving rise to the sequential spontaneous emission of two photons.  The state of the coupled dot-photon system can be calculated within a Wigner-Weisskopf approximation \cite{Scully1997a,Akopian2006a,Avron2008a} when $|\omega_{X_\lambda,0}-\omega_{2X,X_\lambda}|\gg \Gamma_\beta$ \cite{NonLocalFootNote}, where $\beta=\{X_\mathrm{h},X_\mathrm{v},2X\mathrm{h},2X\mathrm{v}\}$, giving the long-time limit
\begin{equation}
 \ket{\psi(\infty)} = \ket{g}\otimes\sum_{\mbf{kq}\lambda} c_{\mbf{kq}\lambda}a_{\mbf{k}\lambda}^\dagger a_{\mbf{q}\lambda}^\dagger\ket{0}.
\end{equation}
The resulting two-photon state is then completely characterized by the coefficients $c_{\mbf{kq}\lambda}$, which have been reported previously in a similar context \cite{Akopian2006a,Meirom2008a,Avron2008a}.  We define the conditional post-selected state of the two photons in the polarization basis by $\rho = \sum_{\lambda\lambda'}\rho_{\lambda\lambda'}\ket{\lambda\lambda}\bra{\lambda'\lambda'}$ with matrix elements
\begin{eqnarray}\label{eq:rho_k}
 \rho_{\lambda\lambda'} &=& \mathrm{Tr}\left( F_{\lambda\lambda'}\ket{\psi(\infty)}\bra{\psi(\infty)}\right)/\eta,\\
F_{\lambda\lambda'} &=& \frac{1}{2}\sum_{\mbf{kq}}\xi_\mbf{k}\xi_\mbf{q}a^\dagger_{\mbf{q}\lambda'}a^\dagger_{\mbf{k}\lambda'}\ket{0}\bra{0}a_{\mbf{k}\lambda}a_{\mbf{q}\lambda}.\label{eq:F}
\end{eqnarray}
Here, $\eta$ is a normalization chosen to enforce $\mathrm{Tr}\rho=1$.  
The coefficients $\xi_\mbf{k}$ simultaneously characterize imperfections in photon detection and postselection due to spectral filtering ($\xi_\mbf{k}=1\,\forall\mbf{k}$ if all emitted photons are detected perfectly).  The form of Eq. (\ref{eq:F}) implicitly assumes that the photon detector bandwidth due to a finite detection time $\tau_d$ is smaller than the relevant features of the biexciton cascade.  In particular, here we assume $\tau_d\gtrsim 1/\delta\omega$.  In the opposite limit, a finite FSS would not limit the entanglement of emitted photons \cite{Bandwidth}.

The degree of bipartite entanglement in polarization for the resulting two-photon state can be characterized by any entanglement measure. Here, we choose the generalized concurrence $C$ \cite{Wootters1998a} which reaches $C=1$ for maximally-entangled states and is $C=0$ for separable states.  For the biexciton cascade, the concurrence is given simply in terms of $\rho_\mathrm{hv}$ (see Eq. (\ref{eq:Concurrence-Efficiency}), below).  The maximum efficiency of collected photon pairs is given by the normalization $\eta$ \cite{Meirom2008a}:
\begin{equation}\label{eq:Concurrence-Efficiency}
 C=2|\rho_\mathrm{hv}|;\;\;\eta = \sum_\lambda\mathrm{Tr}\left(F_{\lambda\lambda}\ket{\psi(\infty)}\bra{\psi(\infty)}\right).
\end{equation}

The concurrence is calculated by converting the wavevector sums in  Eq. \eqref{eq:rho_k} to energy integrals and performing the resulting integrals in the complex plane. For perfect detection ($\xi_\mbf{k} = 1\,\forall\,\mbf{k}$), the concurrence and efficiency evaluate to 
\begin{equation}\label{eq:StandardConcurrence}
C = \frac{\sqrt{\Gamma_{X_\mathrm{h}}\Gamma_{X_\mathrm{v}}\Gamma_{2X_\mathrm{h}}\Gamma_{2X_\mathrm{v}}}}{\Gamma_{2X}\sqrt{\delta\omega^2+\Gamma_X^2}}\simeq\gamma_X\gamma_{2X}\frac{\Gamma_X}{|\delta\omega|};\;\;\eta=1.
\end{equation}
Here,  $ \gamma_{2X}=\sqrt{\Gamma_{2X\mathrm{h}}\Gamma_{2X\mathrm{v}}}/\Gamma_{2X}$ and $\gamma_{X}=\sqrt{\Gamma_{X_\mathrm{h}}\Gamma_{X_\mathrm{v}}}/\Gamma_{X}$ give relative asymmetries in the decay rates (in the following analysis, we will assume $\gamma_X=\gamma_{2X}=1$, a condition approximately realized in experiments \cite{Stevenson2006a,Akopian2006a}). The average exciton and biexciton decay rates are $\Gamma_X = (1/2)\sum_\lambda \Gamma_{X_\lambda}$ and $\Gamma_{2X} = (1/2)\sum_\lambda \Gamma_{2X_\lambda}$, respectively, with individual decay rates (see Fig. \ref{fig:Schematic}) given by Fermi's golden rule: $\Gamma_{X_\lambda}= 2\pi\sum_{\mbf{q}}|f_{\mbf{q}\lambda}|^2\delta\left(\nu_q-\omega_{X_\lambda,0}\right),
\Gamma_{2X_\lambda}= 2\pi\sum_{\mbf{q}}|g_{\mbf{q}\lambda}|^2\delta\left(\nu_q-\omega_{2X,X_\lambda}\right)
$.  When $\gamma_{X}=\gamma_{2X}=1$, Eq. (\ref{eq:StandardConcurrence}) recovers the result reported in Refs. \onlinecite{Avron2008a,Meirom2008a}.  The concurrence is small in the ratio $\Gamma_X/|\delta\omega|$.  This result reflects the fact that the two decay paths given in Fig. \ref{fig:Schematic} are distinguishable in frequency whenever the radiative linewidth is smaller than the frequency difference of emitted photons.  The technology required to apply local electric fields to self-assembled dots is now well-developed, so it is natural to attempt to reduce $\delta\omega$ using a dc field \cite{Gerardot2007a,Reimer2007a}.
An electric field couples to the exciton polarizability, and can be used to reduce $\delta\omega$ by rendering the two-electron wave function more symmetric.  However, a sufficiently strong electric field will simultaneously separate the electron and hole single-particle wave functions, decreasing the radiative linewidth: $\Gamma_X \propto \left|\int d^3r \psi_e(\mbf{r})^*\psi_h(\mbf{r})\right|^2$, making a reduction of the ratio $C\simeq \Gamma_X/|\delta\omega|$ problematic.  
\begin{figure}
	\centering
		\includegraphics[width=0.45\textwidth]{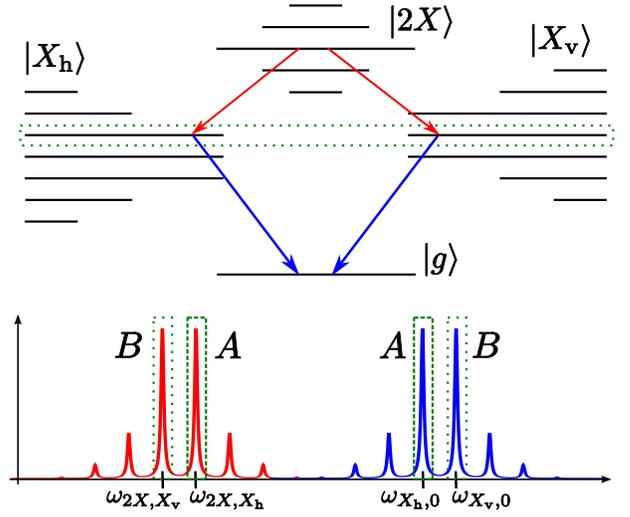}
	\caption{(Color online) Decay cascade and emission spectrum including resolved sidebands from the modulation of the levels and independent spectral filtering for processes $A$ and $B$.}
	\label{fig:Sidebands}
\end{figure}

In the remainder of this Letter, we present a new method for removing which-path information in a biexciton cascade, which yields a finite concurrence in the limit $\Gamma_X/|\delta\omega|\to 0$.   We now introduce an additional time-dependent ac electric (or strain) field, which modulates the exciton and biexciton levels with a periodic drive. In the presence of the ac field, sidebands develop at multiples of the modulation frequency (quasienergy) $\omega$.  By tuning the driving frequency to the FSS ($\omega=\delta\omega)$, we can ``erase'' the which-path information by introducing new decay paths (see Fig. \ref{fig:Sidebands}). We note that the addition of new decay pathways shown in Fig. \ref{fig:Sidebands} does not lead to distinguishability of photon pairs emitted at different times, which may be important for schemes involving joint measurements.  Provided the decay rates and frequencies are fixed, every emitted photon pair will be described by an identical wavepacket, including components from several frequencies.

In general, we account for both the modulation of the energy levels with the replacement $E_j\to E_{j}+\delta E_j\cos(\omega t)$ \emph{and} the modulation of the oscillator strengths, with the replacements $g_{\mbf{k}\lambda}\to G(t)g_{\mbf{k}\lambda}$ and $f_{\mbf{k}\lambda}\to F(t)f_{\mbf{k}\lambda}$, in Eq. \eqref{eq:Hamiltonian}.  The spontaneous emission rates are also modified due to the modulation: $\Gamma_{X,2X}\to\tilde{\Gamma}_{X,2X}$.  It is convenient to go to a generalized rotating frame: $\ket{\tilde{\psi}} = U(t)\ket{\psi}$, with the time-dependent unitary $U(t)=e^{i\int_0^t dt' H_0(t')}$. The Hamiltonian in this frame is:
\begin{equation}\label{eq:HTilde}
 \tilde{H}(t) = \sum_{\mbf{k}\lambda} \tilde{g}_{\mbf{k}\lambda}(t)\sigma_{X_\lambda,2X}a_{\mbf{k}\lambda}^\dagger +\tilde{f}_{\mbf{k}\lambda}(t)\sigma_{0,X_\lambda}a_{\mbf{k}\lambda}^\dagger\\+\mathrm{h.c.},
\end{equation}
where
\begin{multline}
 \tilde{g}_{\mbf{k}\lambda}(t)=\sum_{nm} g_{\mbf{k}\lambda}G_m J_n(\alpha_{2X_\lambda})e^{-i(\omega_{2X,X_\lambda}-\nu_k+[n-m]\omega)t},\\
\tilde{f}_{\mbf{k}\lambda}(t)=\sum_{nm} f_{\mbf{k}\lambda}F_mJ_n(\alpha_{X_\lambda})e^{-i(\omega_{X_\lambda,0}-\nu_k+[n-m]\omega)t},
\end{multline}
with $\alpha_{2X_\lambda}=\alpha_{2X}-\alpha_{X_\lambda}$ and $\alpha_{j}=\delta E_{j}/\omega$.  Here we have exploited the fact that the modulation is periodic, allowing $G(t)$ and $F(t)$ to be written in terms of their discrete Fourier transforms:
 $G(t) = \sum_{m=-\infty}^{\infty} G_m e^{im\omega t},
 F(t) = \sum_{m=-\infty}^{\infty} F_m e^{im\omega t}$.  We have further made use of the Jacobi-Anger expansion: $e^{-i\int_0^t dt'\delta E \cos(\omega t')}=\sum_{n=-\infty}^{\infty} J_n\left(\delta E/\omega\right)e^{-i n\omega t}$, familiar from the theory of photon-assisted tunneling \cite{Tien1963a}.  With the new Hamiltonian (Eq. (\ref{eq:HTilde})) in hand, it is a straightforward but tedious exercise to repeat the steps used to arrive at the concurrence.  However, due to the presence of additional sidebands, it is necessary to introduce the additional condition $|\delta\omega|\gg \Gamma_\beta$ (the resolved-sideband limit) for the simple Wigner-Weisskopf analysis to be valid.  Since this is precisely our regime of interest, no harm is done in making this approximation.

We now give the most relevant specific examples to illustrate the method. First, we consider the case without spectral filtering ($\xi_\mbf{k} = 1$) and we set $\omega=\delta\omega$. Remarkably, to leading order in $\Gamma_\beta/|\delta\omega|$, the contributions from \emph{all} sidebands can be summed analytically.  If the dc component of the oscillator strengths dominates (i.e. F(t)=G(t)=1), we find the decay rates are unchanged $\tilde{\Gamma}_\beta = \Gamma_\beta$, and the concurrence is
\begin{equation}\label{eq:MaxConcurrence}
C = J_1^2(\tilde{\alpha}) +\mathcal{O}\left(\frac{\Gamma_\beta}{\delta\omega}\right);\;\;\eta = 1,
\end{equation}
where $\tilde{\alpha}=\alpha_{X_\mathrm{h}}-\alpha_{X_\mathrm{v}}$ describes the relative modulation amplitude for the two intermediate exciton states.  For this case, the concurrence reaches a maximum $C\simeq 0.3$ when $\tilde{\alpha}\simeq 2$, describing the situation when the FSS approximately vanishes at the peak of the modulation.  Although $C<1$ in this case, several entangled pairs with any finite concurrence can be used to create a single maximally-entangled pair through entanglement purification.  However, purification comes at the cost of additional resources.  For the current case, and with C=0.3, we find that the one-way hashing protocol \cite{Dur2007a} gives a yield of perfect Bell pairs to imperfect pairs of $\simeq 1/15$.


\begin{figure}
	\centering
		\includegraphics[width=0.45\textwidth]{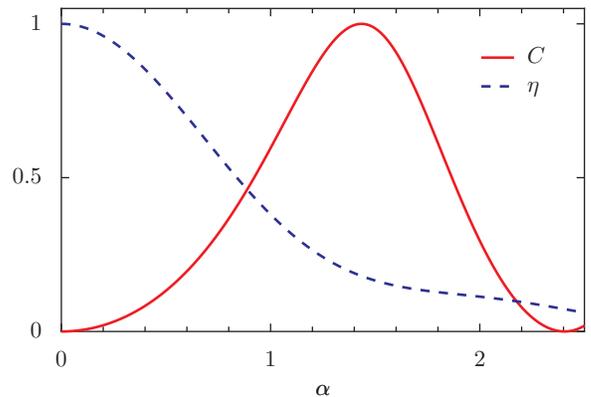}
	\caption{(Color online) Concurrence and efficiency using the optimal filtering method when $\gamma_X=\gamma_{2X}=1$, $\alpha_{X_\mathrm{h}}=\alpha_{X_\mathrm{v}}=\alpha$, and $\alpha_{2X}=2\alpha$.  In the optimal case, where $\alpha_X\simeq1.44$, we have $C=1$, $\eta\simeq0.18$.}
	\label{fig:coneff}
\end{figure}
 Fortunately, it is possible to substantially boost the concurrence in this scheme (to 1 in the ideal case), with a larger efficiency through spectral filtering.  Spectral filtering has been demonstrated previously without level modulation, but has suffered from a relatively low efficiency, limited by the small ratio $\Gamma_j/|\delta\omega|$ \cite{Akopian2006a,Meirom2008a}.  In the presence of sidebands, it is possible to choose the spectral filter advantageously to collect a large fraction of the photons where two sidebands overlap.  Using the filtering scheme shown in  Fig. \ref{fig:Sidebands}, where the photons passing through the windows labeled `$A$' and `$B$' are collected independently with width $w$ satisfying $\Gamma_\beta\ll w \ll \delta\omega$, leads to
\begin{equation}\label{eq:ConcurrenceFilter}
 C=\frac{2J_1^2(\alpha) J_0^2(\alpha)}{\eta}+\mathcal{O}\left(\frac{\Gamma_\beta}{\delta\omega}\right);\;\eta  =  J_0^4(\alpha)+J_1^4(\alpha),
\end{equation}
where we have chosen $\alpha_{X_\mathrm{h}}=\alpha_{X_\mathrm{v}}=\alpha$ and $\alpha_{2X}=2\alpha$ to maximize the concurrence \cite{HalfFreqFootnote}.  The concurrence in Eq. (\ref{eq:ConcurrenceFilter}) reaches a maximum of $C=1$ when $\alpha\simeq 1.4$, leading to a collection efficiency of $\eta\simeq 0.18$ (see Fig. \ref{fig:coneff}; note that $C$ is only sensitive to large fluctuations $\delta\alpha$ of order 1).  Thus, at the expense of a factor of $\sim 5$ reduction in the efficiency of collected entangled pairs, it is still possible to generate maximally-entangled photons with this method, even in the limit $\Gamma_\beta/|\delta\omega|\to0$, and without resorting to entanglement purification.

We have made several implicit assumptions throughout this analysis.  In particular, we assume that there is a reasonable mechanism for modulating the exciton and biexciton levels at GHz frequencies.  This can be achieved, e.g., by coupling the dots to the electric field from a microwave-resonator stripline \cite{Kouwenhoven1994a}, through capacitive coupling to a potential-biased nanomechanical resonator, or by coupling to the strain field due to surface acoustic waves \cite{Sogawa2001a}.  Each of these technologies has been demonstrated in other contexts.  We estimate the electric-field amplitude required to generate maximally-entangled photons is typically $E\sim \hbar\delta\omega/D\sim 10^{3} \mathrm{V}/\mathrm{m}$, where $\hbar\delta\omega\sim 10\,\mu e V$ and $D\sim e\times 10\,\mathrm{nm}$ is the typical exciton dipole moment (dc fields that are larger by a factor of $\sim 10^2$ have been achieved in gated dots \cite{Reimer2008a}). An ac electric field may induce photon-assisted tunneling, but this effect is strongly suppressed whenever $\Gamma_\mathrm{dl}/\tilde{\Gamma}_X\ll 1$, where $\Gamma_\mathrm{dl}$ gives the dot-lead electron tunneling rate.  Fluctuations of the intermediate exciton states during the biexciton cascade can lead to dephasing and a reduced concurrence.  However, if these fluctuations arise from a common bath (e.g., phonons), this is not a fundamental problem in our scheme, since dephasing can be neglected altogether when the two intermediate exciton states are made to couple symmetrically \cite{Hohenester2007a}. This is an advantage of the present scheme over, e.g., time-reordering, which is sensitive to global noise.

We have presented an alternative route to achieving highly-entangled photon pairs rapidly and on demand.  Our suggested method works with typical quantum dots, having a large exciton FSS, and relies on proven technology for local gating of quantum dots, without the explicit requirements of additional laser fields or strong coupling to a cavity.  While other recent proposals for `time-reordering' schemes \cite{Avron2008a,Troiani2008a} suggest that a larger concurrence can be achieved for `on-demand' photons in the presence of a large FSS, we believe the sideband eraser is a promising alternative.  In contrast to time reordering, the sideband eraser requires no large dc electric field to cancel the biexciton binding energy.  Further, maximal entanglement is only achieved in the time-reordering method with the application of a potentially complicated unitary to the photon wavepackets.

Finally, we note that the modulation frequencies involved are typically in the microwave ($\sim 10$ GHz) range, and so we expect direct extensions of this work to allow for an interface between microwave-frequency circuit QED or nano-electromechanical systems (NEMS) and optical-frequency photons.

We thank C. Couteau for initially suggesting that a microwave drive may be used to compensate the FSS, and M. Lukin, A. Imamoglu, and F. K. Wilhelm for useful discussions. We acknowledge support from QuantumWorks, MITACS, MRI, NSERC, an Ontario PDF (WAC), and the CIFAR JFA (WAC and JMG).


\begin{thebibliography}{10}
\newcommand{\enquote}[1]{``#1''}

\bibitem{KnillRaussendorf}
E.~{Knill}, R.~{Laflamme}, and G.~J. {Milburn}, \nat \textbf{409}, 46 (2001); R.~{Raussendorf} and H.~J. {Briegel}, \prl \textbf{86}, 5188 (2001).

\bibitem{BriegelDur}
H.-J. {Briegel}, W.~{D{\"u}r}, J.~I. Cirac, and P.~{Zoller}, \prl \textbf{81}, 5932 (1998); W.~{D{\"u}r}, H.-J. Briegel, J.~I. Cirac, and P.~Zoller, \pra \textbf{59}, 169 (1999).

\bibitem{Benson2000a}
O.~Benson, C.~Santori, M.~Pelton, and Y.~Yamamoto, \prl \textbf{84}, 2513 (2000).

\bibitem{Stevenson2006a}
R.~M. Stevenson, R.~J. Young, P.~Atkinson, K.~Cooper, D.~A. Ritchie, and A.~J. Shields, \nat \textbf{439}, 179 (2006).

\bibitem{Akopian2006a}
N.~{Akopian}, N.~H. Lindner, E.~Poem, Y.~Berlatzky, J.~Avron, D.~Gershoni, B.~D. Gerardot, and P.~M. Petroff, \prl \textbf{96}, 130501 (2006).

\bibitem{Young2009a}
R.~J. {Young}, R.~M. Stevenson, A.~J. Hudson, C.~A. Nicoll, D.~A. Ritchie, and A.~J. Shields, \prl \textbf{102}, 030406 (2009).

\bibitem{GammonBester}
D.~Gammon, E.~S. Snow, B.~V. Shanabrook, D.~S. Katzerl, and D.~Park, \prl \textbf{76}, 3005 (1996); 
G.~Bester, S.~Nair, and A.~Zunger, \prb \textbf{67}, 161306(R) (2003).

\bibitem{Stace2003a}
T.~M. Stace, G.~J. Milburn, and C.~H.~W. Barnes, \prb \textbf{67}, 085317 (2003).

\bibitem{Avron2008a}
J.~E. {Avron}, G.~Bisker, D.~Gershoni, N.~H. Lindner, E.~A. Meirom, and R.~J. Warburton, \prl \textbf{100}, 120501 (2008).

\bibitem{Kowalik2005a}
K.~{Kowalik}, O.~Krebs, A.~{Lema{\^\i}tre}, S.~Laurent, P.~Senellart, P. Voisin, and J.~A. Gaj, App. Phys. Lett. \textbf{86}, 041907 (2005).

\bibitem{Gerardot2007a}
B.~D. {Gerardot}, S.~Seidl, P.~A. Dalgarno, R.~J. Warburton, D.~Granados, J.~M. Garcia, K.~Kowalik, and O.~Krebs, App. Phys. Lett. \textbf{90}, 041101 (2007).

\bibitem{Reimer2008a}
M.~E. {Reimer}, M.~{Korkusi{\'n}ski}, D.~Dalacu, J.~Lefebvre, J.~Lapointe, P.~J. Poole, G.~C. Aers, W.~R. McKinnon, P.~Hawrylak, and R.~L. Williams,  \prb \textbf{78}, 195301 (2008).

\bibitem{Seidl2006a}
S.~{Seidl}, M.~Kroner, A.~{H{\"o}gele}, K.~Karrai, R.~J. Warburton, A.~Badolato, and P.~M. Petroff, App. Phys. Lett. \textbf{88}, 203113 (2006).

\bibitem{Jundt2008a}
G.~{Jundt}, L.~Robledo, A.~{H{\"o}gele}, S.~{F{\"a}lt}, and A.~{Imamo\u{g}lu}, \prl \textbf{100}, 177401 (2008).

\bibitem{Muller2008a}
A.~{Muller}, W.~Fang, J.~Lawall, and G.~S. Solomon, \prl \textbf{101}, 027401 (2008).

\bibitem{JohnePathak}
R.~{Johne}, N.~A. Gippius, G.~Pavlovic, D.~D. Solnyshkov, I.~A. Shelykh, and G.~Malpuech, \prl \textbf{100}, 240404 (2008); P.~K. {Pathak} and S.~{Hughes}, \prb \textbf{79}, 205416 (2009).

\bibitem{Meirom2008a}
E.~A. {Meirom}, N.~H. Lindner, Y.~Berlatzky, E.~Poem, N.~Akopian, J.~E. Avron, and D.~Gershoni, \pra \textbf{77}, 062310 (2008).

\bibitem{Reimer2007a}
M.~E. {Reimer}, M.~{Korkusi{\'n}ski}, J.~Lefebvre, J.~Lapointe, P.~J. Poole, G.~C. Aers, D.~Dalacu, W.~R. McKinnon, S.~{Fr{\'e}d{\'e}rick}, P.~Hawrylak, and R.~L. Williams, arXiv:0706.1075  (2007).

\bibitem{Hohenester2007a}
U. Hohenester, G. Pfanner, and M. Seliger, \prl \textbf{99}, 047402 (2007).

\bibitem{Troiani2008a}
F.~{Troiani} and C.~{Tejedor}, \prb \textbf{78}, 155305 (2008).

\bibitem{Langbein2004a}
W.~{Langbein}, P.~Borri, U.~Woggon, V.~Stavarche, D.~Reuter, and A.~D. Wieck, \prb \textbf{69}, 161301(R) (2004).

\bibitem{Scully1997a}
M.~Scully and M.~Zubairy, \emph{{Quantum Optics}} (Cambridge University Press, 1997).

\bibitem{NonLocalFootNote}
The condition $|\omega_{X_\lambda,0}-\omega_{2X,X_\lambda}|\gg \Gamma_\beta$
  will typically be satisfied due to the additional biexciton
  binding energy, but may be violated in an applied electric field. 

\bibitem{Bandwidth}
C.~{Flindt}, A.~S. {S{\o}rensen}, M.~D. Lukin, and J.~M. Taylor, \prl \textbf{98}, 240501 (2007); R.~M. Stevenson, A.~J. Hudson, A.~J. Bennett, R.~J. Young, C.~A. Nicoll, D.~A. Richie, and A.~J. Shields, \prl \textbf{101}, 170501 (2008).

\bibitem{Wootters1998a}
W.~K. Wootters, \prl \textbf{80}, 2245 (1998).

\bibitem{Tien1963a}
P.~Tien and J.~Gordon, Phys. Rev. \textbf{129}, 647 (1963).

\bibitem{Dur2007a}
W.~{D{\"u}r} and H.~J. {Briegel}, Rep. Progr. Phys. \textbf{70}, 1381 (2007).

\bibitem{HalfFreqFootnote}
It may be difficult to satisfy requirements on both $\alpha_{2X}$ and
  $\alpha_{X_\lambda}$ simultaneously in an experiment. A simpler situation
  arises when we take $\omega=\delta\omega/2$ and introduce a frequency window
  of width $w$ ($\Gamma_{\beta}\ll w\ll \delta\omega$) centered at frequency
  $\bar{\omega}=\omega_{X_\mathrm{h},0}+\delta\omega/2$. In this case, we find
  $C=1$ for $\alpha_{X_\mathrm{h}}=\alpha_{X_\mathrm{v}}=\alpha$ and arbitrary
  $\alpha_{2X},\alpha$. The efficiency in this case is slightly reduced ($\eta$
  reaches a maximum of $\eta=0.11$ for $\alpha =\alpha_{2X}/2= 1.8$), but it
  should be possible to significantly enhance this value by coupling to a
  cavity mode at frequency $\bar{\omega}$.

\bibitem{Kouwenhoven1994a}
L.~P. Kouwenhoven, S.~Jauhar, J.~Orenstein, P.~L. McEuen, Y.~Nagamune, J.~Motohisa, and H.~Sakaki, \prl \textbf{73}, 3443 (1994).

\bibitem{Sogawa2001a}
T.~Sogawa, P.~V. Santos, S.~K. Zhang, S.~Eshlaghi, A.~D. Wieck, and K.~H. Ploog, \prb \textbf{63}, 121307(R) (2001).

\end{thebibliography}

\end{document}